\documentclass[11pt]{IEEEtran}

\usepackage{amssymb}
\usepackage{color}
\usepackage{graphicx}
\usepackage{amsmath}
\usepackage{wasysym}
\usepackage{multirow}
\usepackage{booktabs}
\usepackage{lineno}
\usepackage{soul}
\usepackage[draft]{fixme}
\usepackage{longtable}
\usepackage{supertabular}
\usepackage{authblk}
\usepackage[colon]{natbib}

\newcommand{\delOx}{$\delta{}^{18}\mathrm{O}$ }
\newcommand{\delD}{$\delta\mathrm{D}$ }

\newcommand{\delN}{$\delta{}^{15}\mathrm{N}$ }
\newcommand{\degC}{${}^{\circ}\mathrm{C}$ }
\newcommand{\delAr}{$\delta{}^{40}\mathrm{Ar}$ }

\begin{document}




\title{Water isotope diffusion rates from the NorthGRIP ice core for the last 16,000
years - glaciological and paleoclimatic implications.}


\author[1,2]{V. Gkinis}
\author[1,3]{S. B. Simonsen}
\author[1]{S. L. Buchardt}
\author[2]{J. W. C. White}
\author[1]{B. M. Vinther}



\affil[1]{Centre for Ice and Climate, Niels Bohr Institute, University of Copenhagen,
Juliane Maries Vej 30, DK-2100 Copenhagen, Denmark}
\affil[2]{Institute for Alpine and Arctic Research, University of Colorado, Boulder, 1560 30th Street
Boulder, CO 80303 USA}
\affil[3]{Div. of Geodynamics, DTU space � National Space Institute,
Elektrovej, Build. 327, Kgs. Lyngby, Denmark}

\maketitle

\begin{abstract}

   	\noindent A high resolution (0.05 m) water isotopic record (\delOx) is available
   	from the NorthGRIP ice core. In this study we look into the water isotope
   	diffusion history as estimated by the spectral characteristics of the \delOx
   	time series covering the last 16,000 years.
	The diffusion of water vapor in the porous medium of the firn pack
	attenuates the initial isotopic signal, predominantly having an impact
	on the high frequency components of the power spectrum.
	Higher temperatures induce higher rates of smoothing and thus the signal
	can be used as a firn paleothermometer.
   	We use a water isotope diffusion model coupled to a steady-state densification
   	model in order to infer the temperature signal from the
   	site, assuming the accumulation and strain rate history as estimated using
   	the GICC05 layer counted chronology and a  Dansgaard--Johnsen
	ice flow model.
	The temperature reconstruction accurately
	captures the timing and magnitude of the B\o lling--Aller\o d and Younger Dryas transitions.
	A Holocene climatic optimum is seen between 7 and 9 ky b2k with
	an apparent cooling trend thereafter. Our temperature estimate for the Holocene climatic optimum,
	points to a necessary adjustment of the ice thinning function indicating that the ice flow
	model overestimates past accumulation rates by about 10\% at 8 ky b2k.
	This result, is supported by recent gas isotopic fractionation
	studies proposing a similar reduction for glacial conditions. Finally, the record presents a climatic
	variability over the Holocene spanning millennial and centennial scales with a profound cooling
	occurring at  approximately 4000 years b2k. The new reconstruction technique is able
	to provide past temperature estimates by overcoming the issues apparent in the
	use of the classical \delOx slope method. It can in the same time resolve temperature signals
	at low and high frequencies.

\end{abstract}



%


\section{Introduction}
\label{Intro}

	Polar ice core records provide some of the most detailed views
	of past environmental changes up to 800,000 years before present, in a large part via proxy
	data such as the water isotopic composition and  embedded chemical impurities.
	One of the most important features of ice cores as climate archives, is their
	continuity and the potential for high temporal resolution. The relevance
	of this type of paleoclimate data archive is high within the context of the study of
	Earth's climate system and thus the possibility to predict future climate changes.

	The isotopic signature of polar precipitation, commonly expressed through the $\delta$ notation
	\footnote{
	Isotopic abundances are 	typically reported as deviations of a sample's isotopic ratio
	relative to that of a reference water (e.g. VSMOW) expressed through the $\delta$ notation: $\delta^{i} =
	\displaystyle{\frac{^{i}\textrm{R}_{\textrm{sample}}}{^{i}\textrm{R}_\textrm{{SMOW}}}}
	- 1 \left[\permil \right]$ where $^{2}\textrm{R} = \displaystyle{\frac{^{2}\textrm{H}}{^{1}\textrm{H}}}$
 	and $^{18}\textrm{R} = \displaystyle{\frac{^{18}\textrm{O}}{^{16}\textrm{O}}}$}
	\citep{Epstein1953, Mook2000}
	is related to the temperature gradient between the evaporation and condensation site
	\citep{Dansgaard1964} and has so far been used as a proxy for the temperature
	of the cloud at the time of condensation \citep{Jouzel1984, Jouzel1997, Johnsen2001}.
	Previous studies \citep{Johnsen1989, Lorius1969}
	have reported a linear relationship between the isotopic signal of polar
	precipitation and the temperature at the precipitation site. For Greenland sites and present conditions,
	\cite{Johnsen2001} used a linear relationship between the mean annual
	surface temperature and the mean annual isotopic value of snow described as:
\begin{equation}
\delta{^{18}\mathrm{O}} = 0.67 \cdot T(^{\circ}\mathrm{C}) -13.7~\permil.
\label{eq1_ch4}
\end{equation}
	Assuming that the isotope sensitivity of 0.67 $\permil \mathrm{K}^{-1}$ in Eq.(\ref{eq1_ch4})
	(hereafter ``spatial slope'' $\zeta_s$)
	holds for different climatic regimes (as glacial conditions and inter-stadial events) one can reconstruct
	the temperature history of an ice core site based on the measured \delOx profile.

	The validity of the temperature reconstruction based on $\zeta_s$
	was questioned when studies based on borehole temperature inversion
	\citep{Cuffey1994, Johnsen1995a, DahlJensen1998}
	and gas isotopic fractionation studies \citep{Severinghaus1998,
	Severinghaus1999, Lang1999, Schwander1997, Landais2004} were developed.
	The aforementioned studies drew the following conclusions
	regarding the isotopic thermometer. First, the isotopic slope is not
	constant with time. Second, during glacial conditions the spatial slope
	presents an isotopic sensitivity higher than the sensitivity suggested
	by the borehole inversion and gas isotopic studies.
	As a result,
	reconstructions of past temperatures based on the water isotope signal,
	should assess the sensitivity of the \delOx to temperature for
	different climatic regimes, thus inferring the ``temporal slope''
	$\zeta_t$.

	The main reasons that drive the fluctuations of $\zeta_t$ with time are not precisely determined yet.
	Changes of the vapor source location and temperature, the possible presence of an atmospheric inversion
	over the precipitation site, microphysical cloud processes affecting the in--cloud phase changes,
	as well as effects related to the seasonality of the precipitation, have previously been proposed
	as possible causes of this behavior \citep{Jouzel1997}. Furthermore, \cite{Vinther2009} showed how
	changes of the ice sheet geometry affect the elevation of ice core sites and can create isotopic artifacts,
	thus ``masking'' the Holocene climatic optimum signal from ice core records.
	The complications emerging from the variable nature of $\zeta_t$, have possibly also resulted in the
	isotopic signal pointing towards a climatically stable Holocene epoch. A very low signal
	to noise ratio is observed in the \delOx signal of almost every deep ice core from the Greenland
	ice divide. This  observation contradicts
	with studies that indicate that the Holocene epoch has  not been as stable as previously thought
	and likely  significant climatic changes have occurred at high latitude sites \citep{Denton1973, Bond2001}.
	More recently, and based on combined  \delN and \delAr gas analysis on the GISP2 ice core
	\cite{Kobashi2011} suggested temperature variations up to 3 K during the last 4,000 years.
	The impacts of these changes are believed to have been significant for the evolution of past
	human civilizations \citep{Dalfes1997, deMenocal2001, Weiss2001}.

	An alternative way to extract temperature information from the water isotopic signal,
	is to look into its spectral properties.
	Inherently noisy at the beginning  and affected by
	a number of post depositional effects, the isotopic signal experiences
	attenuation via a molecular diffusion process during the stage of firn densification.
	The diffusion process eliminates the power of the high frequency components of the
	spectrum and often results in a more common \delOx signal between ice cores whose upper
	parts show a very low cross-correlation of their \delOx time series \citep{Fisher1985}.
	Information on this
	process is contained in the spectral characteristics of the \delOx signal.
	We work here
	along the lines of previously published work by \cite{Johnsen1977}, \cite{Johnsen2000}
	and \cite{Simonsen2011} presenting a technique that is based on the spectral analysis of
	the \delOx signal only, thus not requiring dual isotope measurements of both \delOx and \delD
	as in \cite{Simonsen2011}.
	An estimate of the densification process is vital for this type of study and can be carried out in
	various ways \citep{Herron1980, Alley1982, Schwander1997, Goujon2003},
	with each approach posing its own uncertainty levels, implementation characteristics
	and challenges.
	The technique yields actual firn temperatures and thus overcomes many of the issues of the
	traditional \delOx thermometer based on the ``temporal slope'' $\zeta_t$.


\section{The water isotope diffusion - densification model}
\label{sec:model}

	We will outline here the theory of water isotope diffusion in firn and ice.
	Almost exclusively, the formulations used here are based on the work presented in
	\cite{Johnsen1977} and \cite{Johnsen2000} and the experimental work of \cite{Jean-Baptiste1998}
	and \cite{VanderWel2011}.

\subsection{Firn diffusion}

	The diffusion of water isotopes in firn is a process that occurs after the deposition of
	precipitation. It is a molecular exchange process taking place in the vapor phase
	and it is driven by isotopic gradients existing along the firn column. As the densification
	of firn continues, the diffusion process slows down until it ceases at pore close off.
	Assuming a coordinate system fixed on a sinking layer of firn, the process can be described mathematically
	by  Fick's second law accounting for layer thinning as:
\begin{equation}
\frac{\partial \delta}{\partial t} = D \left( t \right) \frac{\partial^2 \delta}{\partial z^2} -
\dot{\varepsilon}_z \left( t \right) z ~\frac{\partial \delta}{\partial z} \enspace .
\label{eq2.1}
\end{equation}
	Here, $D \left( t \right)$ is the diffusivity coefficient and $\dot{\varepsilon}_z\left(t\right)$
	the vertical strain rate. Use of Fourier integrals
	yields the solution to Eq.(\ref{eq2.1}) as the convolution of the initial isotopic signal
	$\delta \left(z, 0\right)$ with a Gaussian filter of standard deviation $\sigma$:
\begin{equation}
\mathcal{G} = \frac{1}{\sigma \sqrt{2\pi}} \, e^{\frac{-z^2}{2 \sigma^2}} \enspace ,
\label{eq2.2}
\end{equation}
	The isotopic signal $\delta \left( z, t \right)$ will then be given by
\begin{equation}
\delta \left( z, t \right) = \mathcal{S} \left( t \right) \frac{1}{\sigma \sqrt{2 \pi}}
\int_{-\infty}^{+\infty} \delta \left( z, 0 \right) \exp \left\{ \frac{-\left(z-u \right)^2}
{2 \sigma^2}\right\} \,\mathrm{d}u \enspace .
\label{eq2.3}
\end{equation}
	Here $\mathcal{S} \left( t \right)$ is the total thinning the layer has experienced during
	the time interval $t=0 \rightarrow t=t'$ due to the ice flow and equal to:
\begin{equation}
\mathcal{S} \left( t' \right) = e^{\int_0^{t'} \dot{\varepsilon}_z \left( t \right) \mathrm{d}t} \enspace .
\label{eq2.4}
\end{equation}
	The standard deviation term  $\sigma$ of the Gaussian filter, commonly referred to as diffusion length,
	represents the mean displacement of a water molecule along the $z$-axis and can be calculated
	as \citep{Johnsen1977}:
\begin{equation}
\frac{\mathrm{d}\sigma^2}{\mathrm{d}t} - 2\,\dot{\varepsilon}_z\!\left( t \right) \sigma^2 =
2D\!\left( t \right) \enspace .
\label{eq2.5}
\end{equation}
		In the firn the simple strain rate can be assumed
\begin{equation}
\dot{\varepsilon}_z \left(
	t \right) = -\frac{\mathrm{d\rho}}{\,\,\,\mathrm{d t}}\,\frac{\,1\,}{\,\rho\,}.
\label{eq2.6}
\end{equation}
	Combining Eq.(\ref{eq2.5}) and Eq.(\ref{eq2.6}) and substituting the independent
	variable t with the density $\rho$, we can calculate the quantity $\sigma^2$
	provided an expression of the diffusivity $D\left(\rho\right)$ (see SOM), a density
	profile for the site and its adjoint age are estimated.
	The expression we finally get for the diffusion length is (see SOM for a more detailed derivation):
\begin{equation}
\sigma^2 \left( \rho \right) = \frac{\,1\,}{\rho^2}\int_{\rho_o}^{\rho}2\rho^2
{\left( \frac{\mathrm{d}\rho}{\mathrm{d}t}\right)}^{-1}\! D \!\left( \rho \right) \,\mathrm{d}\rho ,
\label{eq2.7}
\end{equation}
	where with $\rho_o$ we symbolize the surface density.
	In this study we used
	the empirical steady-state densification model by \cite{Herron1980} (H--L model hereafter), according to which:
\begin{equation}
\frac{\mathrm{d}\rho(z)}{\mathrm{d}t} = K(T)A^{\vartheta} \left(\rho_{\mathrm{ice}} - \rho \left(z\right) \right).
\label{eqhl}
\end{equation}
	Here $K(T)$ is a temperature dependent Arrhenius--type densification rate coefficient, $A$ is the annual
	accumulation rate (here in $\mathrm{kgm}^{-2}\mathrm{yr}^{-1}$) and $\vartheta$
	is a factor that determines the effect of the accumulation rate
	during the different stages of the densification. For a better insight on those parameters
	the reader is referred to \cite{Herron1980}.

	The mathematical treatment of the diffusion process presented in this section
	does not take into account possible effects of barometric wind pumping in the top firn layer.
	Pressure gradients caused by wind flowing over the surface topography
	\citep{Colbeck1989, Waddington2002}   or by diurnal temperature fluctuations
	\citep{Hindmarsh1998} can cause Darcian vapor fluxes penetrating the firn.
	These fluxes can  alter the isotopic composition of the upper firn.
	The importance of these effects depends on parameters like temperature, accumulation,
	wind and surface topography. Considering the relatively high accumulation rates
	at the NorthGRIP site, it is likely that the surface snow is not exposed to the overlying
	vapor at times sufficiently long to result in isotopic changes that will have a major
	impact in our reconstruction. We acknowledge that diffusion models that take these
	effects into account are likely to give a more accurate reconstruction of past temperatures.
	For the very low accumulation sites of the East Antarctic plateau these affects can
	be significant and should be taken into account.

\subsection{Diffusion below the close--off depth}
	At the close--off depth, the process of diffusion in the vapor phase ceases.
	After $\rho \geq \rho_{\mathrm{ice}}$, self diffusion takes place in the solid phase.
	For the temperature dependence
	of the diffusivity we use an Arrhenius type  equation as \citep{Ramseier1967, Johnsen2000}:
\begin{equation}
D_{ice} = 9.2 \cdot 10^{-4} \cdot \exp \left(- \frac{7186}{T} \right) \mathrm{ m^2 s^{-1}} .
\label{eq2.14}
\end{equation}
	In order to calculate the quantity $\sigma_{ice}$ we start from Eq.(\ref{eq2.5})
	and use $D_{ice}$ as the diffusivity parameter,
\begin{equation}
\frac{\mathrm{d}\sigma_{ice}^2}{\mathrm{d}t} - 2\,\dot{\varepsilon}_z\!\left( t \right) \sigma_{ice}^2 =
2D_{ice}\!\left( t \right).
\label{eq2.15}
\end{equation}
	The solution for Eq.(\ref{eq2.15}) yields (see SOM):
\begin{equation}
\sigma_{ice}^2 \!\left( t' \right) =
S\! \left( t' \right)^2 \,\int_0^{t'} 2 D_{ice} \!\left( t \right) S \!\left( t \right) ^{-2} \,\mathrm{d} t
\label{eq2.16}
\end{equation}
	For a further discussion on the
	ice diffusion coefficient the reader is referred to the SOM.

\subsection{The total diffusion signal at depth $z_i$}
	If with $\sigma^2_{\mathrm{firn}}$ we symbolize the firn diffusion length
   	as estimated by performing the integration in Eq.(\ref{eq2.7}) from the
   	surface density $\rho_o$ to the close--off density $\rho_{\mathrm{co}}$ and
   	expressed in ice equivalent length, then the total diffusion length at a depth $z_i$ will be:
\begin{equation}
\sigma^2_i \left( z \right) = \left[S(z)\sigma_{\mathrm{firn}}(z)\right]^2 + \sigma^2_{\mathrm{ice}}(z)
\label{eq2.17}
\end{equation}

	In this study, for the estimation of $\sigma^2_{\mathrm{ice}}$ we use the temperature
	profile from the borehole of the NorthGRIP core \citep{DahlJensen2003} and assume
	a steady state condition.

\begin{figure*}
\vspace*{2mm}
\center
\includegraphics[width = 80mm]{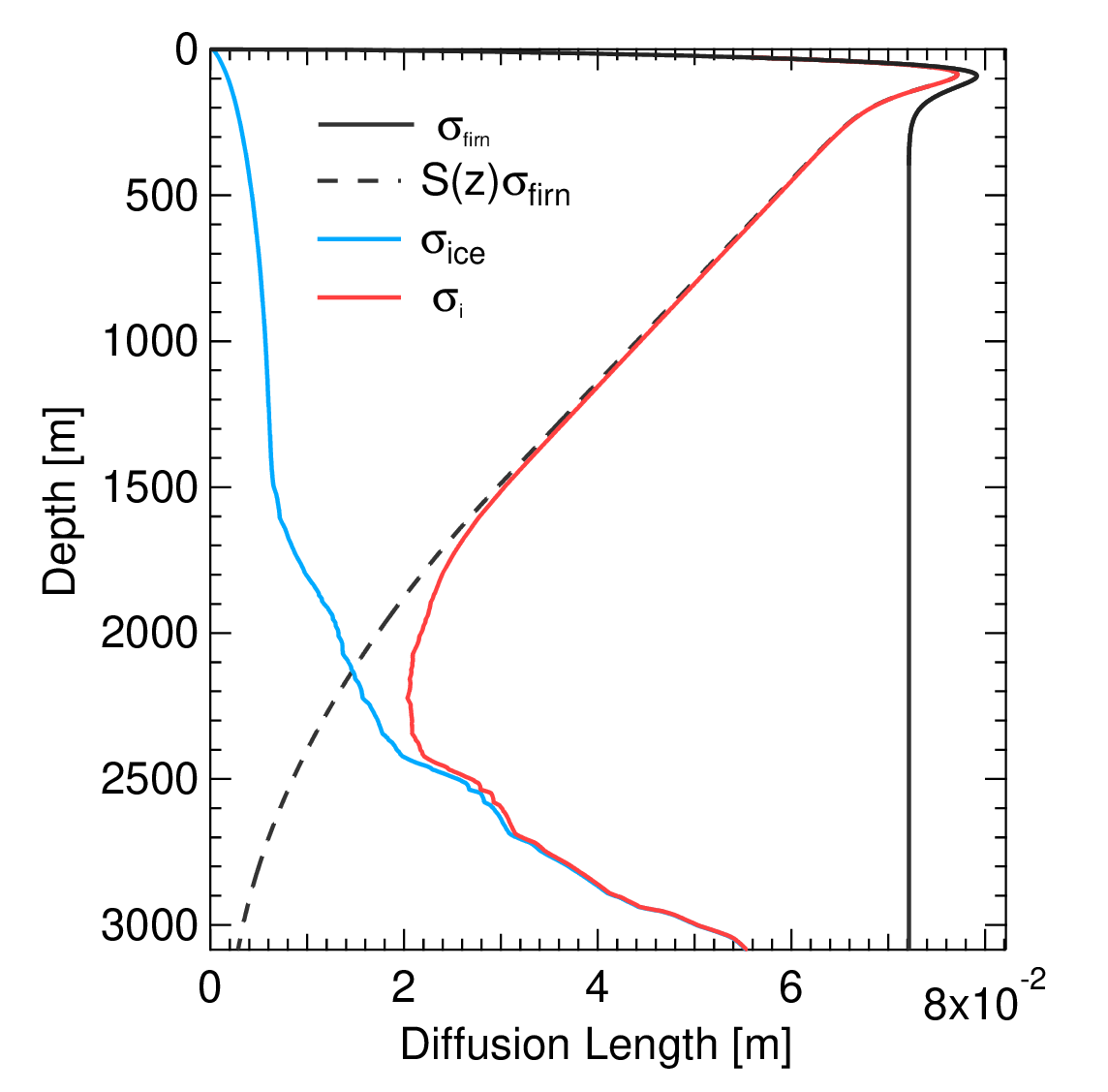}
\caption{Vertical profiles of $\sigma_{\mathrm{firn}}$, $S(z)\sigma_{\mathrm{firn}}$, $\sigma_{\mathrm{ice}}$ and $\sigma_i$
	For the calculation of
	$\sigma_{\mathrm{firn}}$ the parameters we used for the H--L model were:
	$P = 0.7$ Atm, $\rho_0 = 330 \mathrm{\;kgm^{-3}}$,
	$\rho_{\mathrm{CO}} = 804.3 \;\mathrm{kgm^{-3}}$, $T = 242.15$ K, and
	$A = 0.2 \;\mathrm{myr}^{-1}$ ice equivalent.}
\label{Fig5}
\end{figure*}

	In Fig. \ref{Fig5} the
	quantities $\sigma_i$, $\sigma_{\mathrm{ice}}$, $\sigma_{\mathrm{firn}}$ and
	$S(z)\sigma_{\mathrm{firn}}$ are illustrated for typical modern day conditions at NorthGRIP.
	These parameters are only
	approximately representative of the modern NorthGRIP conditions and
	chosen as such, solely for the purpose of illustration.

\section[Estimation of the diffusion length based on high resolution datasets]
{Estimation of the diffusion length based on high resolution \delOx datasets}
\label{spec_est}

	The transfer function for the Gaussian filter in Eq. (\ref{eq2.2}) will be given by
	its Fourier transform, which is itself a~Gaussian and  equal to

\begin{equation}
\label{ftrans_gaussian}
\mathfrak{F}
[ \mathcal{G}_{\rm cfa} (z) ] =
\hat{\mathcal{G}}_{\rm cfa} = {e}^{\frac{-k^2 \sigma^2}{2}} {} \enspace ,
\end{equation}
	where $k = 2\pi / \Delta$ and $\Delta$ is the sampling resolution of the \delOx time series \citep{Abramowitz1964}.
	Harmonics with an initial amplitude $\Gamma_0$ and
	wavenumber $k$ will be attenuated to a~final amplitude $\Gamma_s$ as described in Eq.(\ref{attenuation}):
\begin{equation}
\label{attenuation}
\Gamma_s = \Gamma_0 {e}^{\frac{-k^2 \sigma^2}{2}} {}
\end{equation}
	Based on the latter, one can describe the behavior of the power spectrum of the isotopic time
	series that has been subjected to the firn diffusion process, assuming that at the time of
	deposition the power spectral density shows a white noise behavior. For a time series that
	is sampled at resolution equal to $\Delta$ and presenting a noise level described
	by $  \eta \left( k \right) $,   the power spectral density will be described as:
\begin{align}
\label{powersd}
P_s &=  P_{\sigma}  + {\vert \hat{\eta} \left( k \right) \vert} ^{2} \;\; \\\mathrm{where} \;\;
\label{powersd1} P_{\sigma} = P_0 {e}^{-k^2 \sigma^2}&, \;\;k=2\pi f  \;\;
\mathrm{and} \;\;  f \in \left[ 0, \; \frac{1}{2\Delta}  \right]
\end{align}
	From Eq.(\ref{powersd}, \ref{powersd1}) one can see that the estimation of the diffusion length
	$\sigma$ benefits from time series sampled at a high resolution $\Delta$ and presenting
	a low noise level $  \eta \left( k \right) $.

\begin{figure*}
\vspace*{2mm}
\center
\includegraphics[width=110mm]{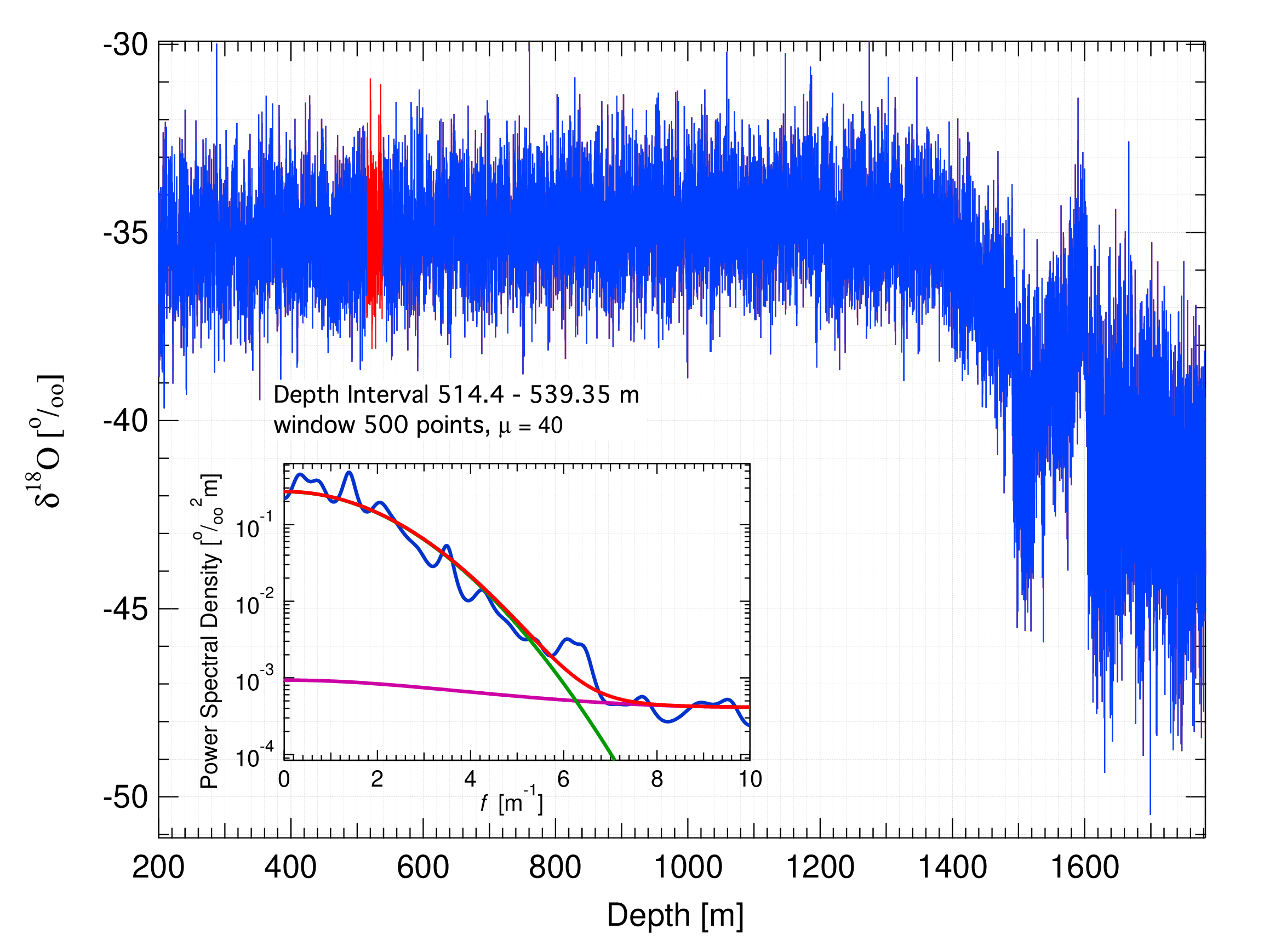}
\caption{NorthGRIP high resolution (0.05 m) \delOx record. The subplot is an example of a power
spectral density estimation with the MEM (blue curve). The length of the window  is 500 points
and the order of the AR model for the spectral estimation is $\mu = 40$. The model for the spectral density $P_s$ is presented
in red, $|\hat{\eta}(k )|^2$ is shown in purple and $P_{\sigma}$ in green.}
\label{Fig6}
\end{figure*}

	We obtain an estimate $\hat{P}_s$ of the power spectral density
	density $P_s$ by using Burg's spectral estimation method and
	assuming a $\mu$-order autoregressive (AR-$\mu$)
	process \citep{Kay1981, Hayes1996}.
	On a sliding window with a length of 500 points (25 m),
	we apply the algorithm described in \cite{Andersen1974}.
	We calculate a value for the diffusion length $\sigma^2$, by minimizing the misfit
	 between $P_s$ and $\hat{P_s}$ in the least squares sense.
	An example of an estimated spectrum and the inferred model is given in Fig. \ref{Fig6}.

	The key assumption underpinning the diffusion length estimation is that variations in the frequency
	content of the primary record (i.e., the \delOx of snowfall before diffusive smoothing)
	through time are small compared to the imprint of isotopic diffusion itself.
	This assumption is supported by synthetic data tests described in section 6 of the SOM
	where additional information on the spectral and $\sigma^2$ estimation can also be found.

	It is worth noting that the spectral estimation does not require that the
	\delOx record is corrected for changes in elevation or the \delOx content
	of sea water.

\section{Inferring temperatures from the estimated diffusion lengths}
\label{sectemp}

	Let us assume that  the total diffusion length value $\widehat{\sigma}^2_i$ at depth $z_i$
	is estimated from the power spectral density and
	a combination of firn densification and flow model parameters as well as
	an estimate of past accumulation rates is known.
	The first step towards obtaining a temperature
	$T_i$  is to account for the artifactual diffusion that occurs due to the finite sampling
	scheme. We require that  the transfer function of a rectangular filter with
	width equal to $\Delta$, is equal to the transfer function of a Gaussian filter with
	diffusion length $\sigma^2_{dis}$. This results in:
\begin{equation}
\sigma^2_{dis} = \frac{2\Delta^2}{\pi^2}\log{\left(\frac{\pi}{2}\right)}.
\label{aliasing}
\end{equation}
	Then, the sampling diffusion length is subtracted from $\hat{\sigma}^2_i$ in the Gaussian sense as:
\begin{equation}
\sigma^2_i = \widehat{\sigma}^2_i  -  \sigma^2_\text{dis}.
\label{sam_corr}
\end{equation}
	Combining Eq.(\ref{eq2.17}, \ref{sam_corr}) we get
\begin{equation}
\sigma^2_\text{firn} = \frac{\widehat{\sigma}_i^2 - \sigma^2_\text{dis} - \sigma^2_\text{ice}}{{S(z)}^{2}}.
\label{firn_data}
\end{equation}
	The result of this calculation describes the data based diffusion
	length of the layer under consideration at the close--off depth in m  ice eq.

\begin{figure*}
\vspace*{2mm}
\center
\includegraphics[width= 120mm]{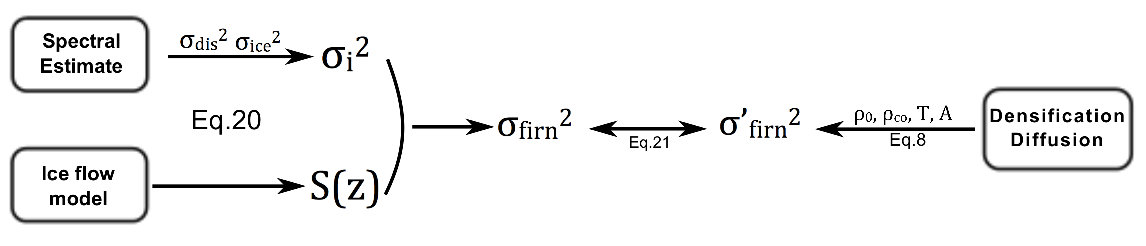}
\caption{Block diagram of the computation scheme for the temperature reconstruction.
Data-based estimates of the $\sigma_{\mathrm{firn}}^2$ parameter are obtained
from the spectral estimation procedure and then  corrected for discrete sampling, ice diffusion and
ice flow thinning effects.
A model based estimate of the diffusion length $\sigma'^2_\mathrm{firn}$ using an isothermal
firn layer of temperature $T$ and accumulation $A$ is obtained using the densification/diffusion calculation.
Finally we compute the roots of the term $\sigma'^2_\mathrm{firn} - \sigma_{\mathrm{firn}}^2$ using $T$ as the independent
variable.}
\label{block_diagram}
\end{figure*}

	A  model based calculation of the diffusion length at the close--off depth
	can be obtained that will allow the calculation of the temperature $T$.
	Central to this
	calculation is Eq.(\ref{eq2.7}) where $\rho = \rho_{\mathrm{co}}$ and the integration
	takes place from the surface density $\rho_o$ to the close--off density $\rho_{\mathrm{co}}$.
	The primed notation
	is used for the model based diffusion length estimate as $\sigma'^2_\mathrm{firn}$.
	Additional to the temperature $T$, the model also requires a value for the accumulation rate parameter $A$.
	Our accumulation rate estimates are based on the combined information from the ice flow model and
	the estimate of the annual layer thickness, available from the GICC05 chronology (see section \ref{accum_dating}).
	The temperature estimation uses the Newton-Raphson method in order to find the roots of the term:
\begin{equation}
\underbrace{\frac{\rho_{\mathrm{co}}}{\rho_{\mathrm{ice}}}\;\sigma^2\left(\rho = \rho_{\mathrm{co}}, T( z ), A ( z ) \right)}_{\sigma'^2_\mathrm{firn}} - \sigma^2_{\mathrm{firn}}.
\label{newton_raphson}
\end{equation}
	The computation is summarized in the block diagram in Fig. \ref{block_diagram}.

\section{The NorthGRIP diffusion length based temperature history}

\subsection{The high resolution \delOx NorthGRIP record}

	The NorthGRIP ice core was drilled at $75.10\;{}^{\circ}\mathrm{N}$, $42.32\; {}^{\circ} \mathrm{W}$ at an elevation
	of 2,917 m, to a depth of 3,085  m and extending back to 123,000 years
	\citep{NGRIPmembers2004}.
	A high resolution record of \delOx is available from this ice core with a resolution of $\Delta = 0.05$ m.
	Stable isotope \delOx analysis were performed at the stable isotope laboratory of the
	University of Copenhagen using a SIRA10 isotope ratio mass spectrometer with an adjacent
	$\mathrm{CO}_{2} - \mathrm{H}_2 \mathrm{O}$ isotopic equilibration system.
	The analytical uncertainty of the system is  0.06 $\permil$.
	This number refers to the analytical performance of the mass spectrometry
	system alone and does not necessarily describe the variance of the \delOx
	time series.
	The latter is largely affected by the accumulation patterns at the site
	as well as effects occurring after deposition of the snow \citep{Fisher1985}
	The high resolution \delOx record down to 1700 m is presented in Fig. \ref{Fig6}.

\subsection{Dating - accumulation and strain rate history}
\label{accum_dating}

\begin{figure*}
\vspace*{2mm}
\center
\includegraphics[width=80mm]{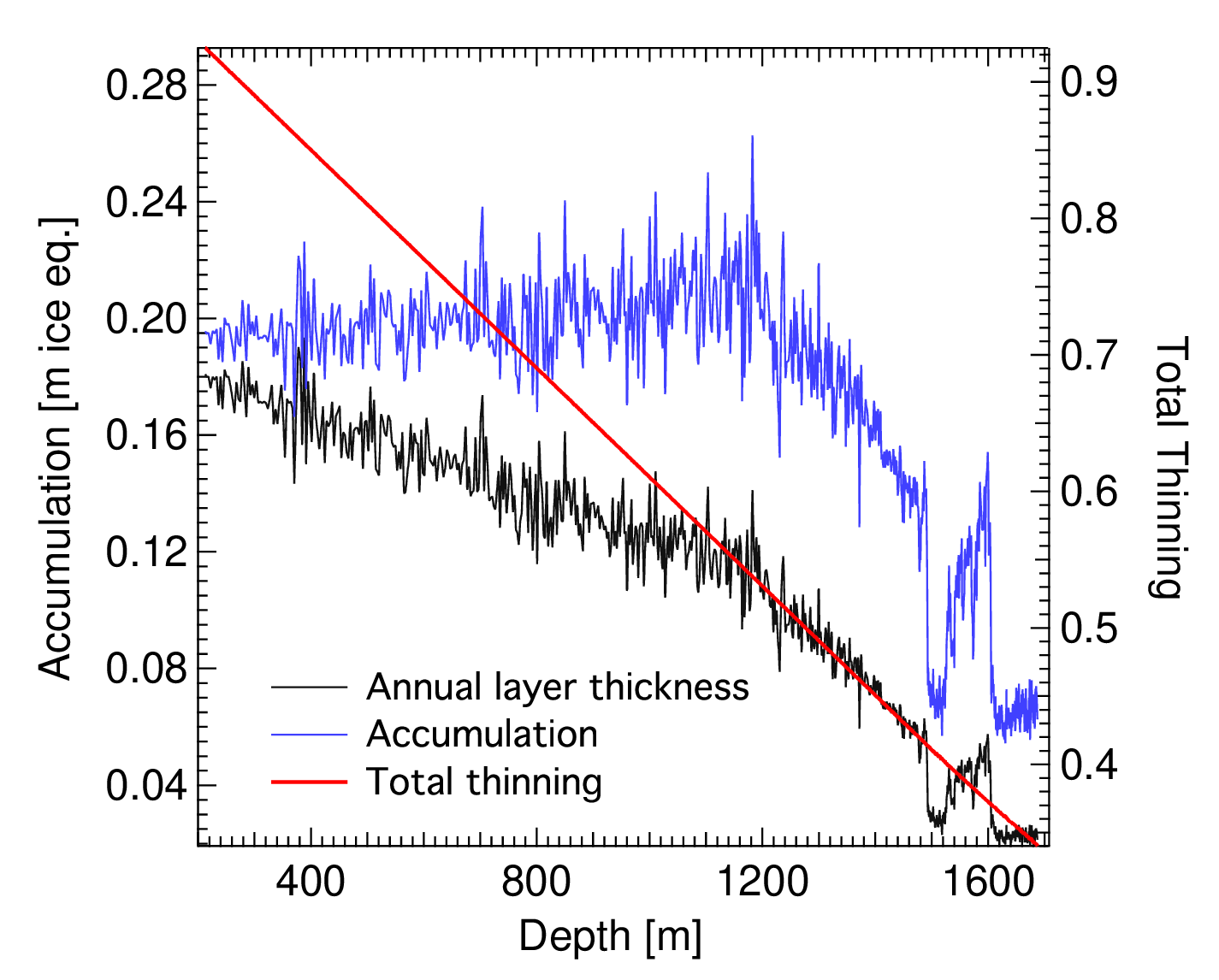}
\caption{Annual layer thickness, total thinning and accumulation rate for NorthGRIP based on the GICC05 chronology and
a 1--D Dansgaard--Johnsen ice flow model}
\label{Fig10}
\end{figure*}

	The NorthGRIP ice core has been dated by manual counting of annual layers down to 60 ka,
	as a part of the Greenland Ice Core Chronology 2005 (GICC05)
	\citep{Vinther2006, Rasmussen2006, Andersen2006, Svensson2008}.  In this study,
	a combination of the GICC05 estimated annual layer thickness $\lambda(z)$ and a
	thinning function $S(z)$ inferred by
	an ice flow model provides an accumulation rate history $A(z)$ for the site. We use a
	Dansgaard-Johnsen (hereafter D--J model) type 1--D ice flow model
	\citep{Dansgaard1969} with basal melting and sliding.
	The model is inverted in a Monte--Carlo fashion allowing estimation of the ice flow parameters
	and constrained by measured depth--age horizons.
	The values inferred from the model for the present day
	accumulation rate, basal melting and the kink height are
	$A_{\mathrm{modern}} = 0.189 \;\mathrm{m\,y}^{-1}$ (ice equiv.),
	$b = 7.5\; 10^{-3} \; \mathrm{m\,y}^{-1}$ and
	$\mathcal{H} = 1620 \;\mathrm{m}$ respectively. In Fig. \ref{Fig10},
	we present the annual layer thickness together with the estimated
	thinning function and the accumulation rate. Note how the $S(z)$ varies linearly with depth.
	This is a result of the D--J flow model that requires the horizontal
	velocity to be constant from the surface until the kink height
	of $\mathcal{H} = 1620 \mathrm{m}$ (1465 m depth).

\subsection{The diffusion length record for NorthGRIP}
	For the calculation of the diffusion length profile we follow the power spectral estimation
	approach as described in section \ref{spec_est} and section 4 in the SOM.
	In order to investigate the stability of the spectral estimation technique we vary the value
	of the order $\mu$ of the AR filter in the interval $\left[40,80\right]$.
	This results in 41 estimates of the diffusion length for every depth, providing an estimate of the
	stability of the power spectral estimation  as well as the variance of the diffusion length estimation.

\begin{figure*}
\vspace*{2mm}
\center
\includegraphics[width=80mm]{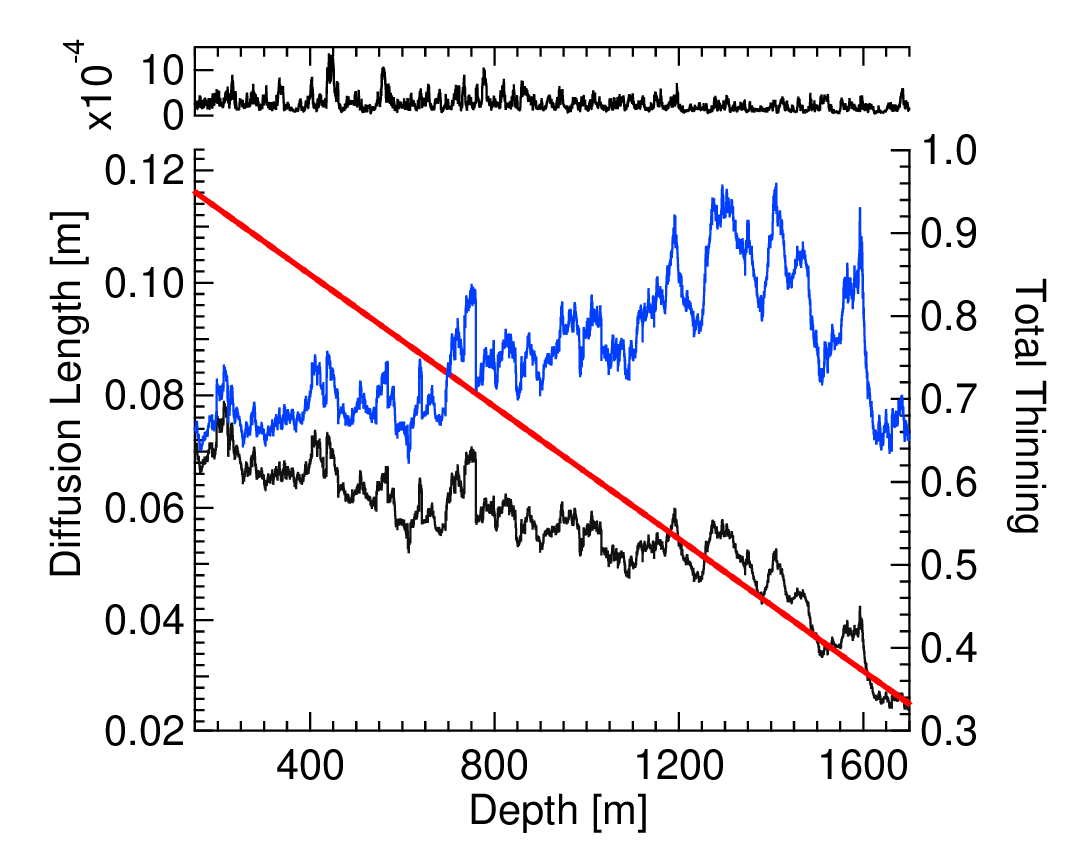}
\caption{Results of the diffusion length estimation.
Bottom plot: raw diffusion length values
$\sigma^2_{\mathrm{i}}(z)$ (bottom black curve)
and $\sigma^2_{\mathrm{firn}}(z)$ (blue curve) after correcting for total thinning (red curve)
and $\sigma^2_{\mathrm{ice}}$
Top plot: Standard deviation
of the 41 estimates of the diffusion length for every depth in m ice eq.}
\label{Fig8}
\end{figure*}

	In Fig. \ref{Fig8},  we present the result of the this estimation representing the value of
	$\widehat{\sigma}_i^2$ together with its adjoint variance (top panel in Fig. \ref{Fig8}).
	We also show the result of the discrete sampling, ice flow thinning and ice
	diffusion correction representing the value of $\sigma^2_\text{firn}$. This is
	the data-based diffusion length value we use in order to infer the temperature
	of the record using the computation scheme described in section \ref{sectemp}
	and illustrated in Fig. \ref{block_diagram}. Based on a sensitivity study described in
	detail in section 5, 6 and 8 of the SOM we estimate the uncertainty of the reconstruction to
	be approximately 2.7 K $(1\sigma)$.

\section{Discussion}
\subsection{General picture}

	Based on a diffusion length estimation for the last 16,000 years we reconstructed a
	temperature history for the NorthGRIP site. In Fig. \ref{Fig16} we present the temperature reconstruction
	after applying two sharp low--pass filters with a cut--offs at 200 and 1000 y.
	By perturbing the diffusivity model parameters with varying atmospheric pressure values,
	we have found our temperature inversion to be insensitive to ice sheet elevation changes
	within the range estimated in \cite{Vinther2009}.
	The estimation of the temperature signal presented here does not include any lapse rate corrections.
	The temperature at every point in time is the temperature at the surface of the NorthGRIP
	site.

	The reconstruction stops
	at 225y b2k, a time that corresponds approximately to the present close--off depth.
	The diffusion based reconstruction, results in a mean  temperature between 250 and 350 y b2k
	of 241.4 K (-31.7 C).
	This is a reasonable estimate within the
	prescribed error bars, considering the modern temperature at the NorthGRIP site of
	$-31.5$\degC \citep{NGRIPmembers2004}. Additionally the temperature history presents
	an obvious Holocene optimum at $\approx 8$ky b2k. However, a temperature gradient of 10 K
	between the Holocene optimum and the present time is not supported by any previous study
	and is likely unrealistic. We assess this issue in more detail in section \ref{thinning_section}

	The record shows two prominent transitions from colder to warmer climatic conditions
	at $\approx$ 14.7ky and 11.7ky respectively resembling the B\o lling--Aller\o d
	(BA) and Younger Dryas (YD) oscillations. Based on our reconstruction,
	the amplitudes of those events are found to be $10.7 \pm 2.7 \mathrm{K}$
	and $7 \pm 2.7 \;\mathrm{K}$ respectively.
	These results are in line with $\delta^{15}\mathrm{N}$ studies from the summit of Greenland as presented in
	\cite{Severinghaus1998, Grachev2005} and \cite{Severinghaus1999}.
	In our reconstruction, the absolute temperature right before the beginning of the warming of the YD termination
	is calculated to be $232 \pm 2.7 \;\mathrm{K}$, a temperature that is  higher than the
	suggested 227 K in \cite{Severinghaus1998}.

	When calculated using the ``classical'' isotope thermometer and assuming a
	constant sensitivity of $0.67 \;\permil \mathrm{K}^{-1}$, the magnitudes
	of these oscillations are underestimated by almost a factor of two.  To overcome
	this problem, \cite{Johnsen1995a} tuned the isotope sensitivity using the low resolution
	borehole inferred temperature history \citep{Cuffey1995, DahlJensen1998} and assuming
	a quadratic relationship between temperature and the \delOx signal.
	With this in mind and
	considering that our result is based on the \delOx signal itself,
	we find our conclusions to be relevant within  the discussion regarding  the overall validity
	and eventually the importance of the isotope thermometer.

	At approximately 8.5 ky b2k we observe a strong cooling event ($\approx 5$ K) that lasts about 500 years.
	Similar paleoclimatic signals indicating widespread cooler/drier conditions have been reported
	from regions of the North Atlantic \citep{Andrews1999, Risebrobakken2003, Rohling2005} as well as
	monsoon areas \citep{Gasse2000, Haug2001, Gupta2003, Wang2005}. The magnitude and age of the event
	resemble those of the 8.2k event observed in the \delOx signal of Greenland ice cores
	\citep{Alley1997}. Even though the 8.2k event could possibly be seen as part of a more general climate instability
	\citep{Ellison2006}, it would be speculative to propose that the  North Atlantic freshening mechanism that likely
	triggered the 8.2k event is related to the longer cooling event we observe here.

\subsection{Proposed thinning function correction}
\label{thinning_section}

	Considering the above features of our temperature reconstruction, we can
	comment that apart from the Holocene
	optimum discrepancy and warm glacial, the general characteristics of the signal
	are in line with the synthesized picture we have
	for the Greenland summit temperature history over the last 16ky years.
	This picture is based on a variety of
	methods and proxies, most prominently the gas isotopic fractionation \delN studies and the borehole inversion
	technique.
	Additionally, our technique seems to overcome some of the problems of the classical interpretation
	of the \delOx signal, caused by the variable nature of the isotope temperature sensitivity with time.

	With that in mind, we hereby propose that the observed Holocene optimum discrepancy is due to
	an inaccurate correction for the ice thinning $S\left( z \right)$ as described in Eq.(\ref{eq2.17}) and
	illustrated in Fig. \ref{Fig8}.
	In order to account for this effect we correct $\sigma_i^2$ with a modified
	thinning function $S'\left( z \right)$. We tune $S'\left( z \right)$ such that the obtained temperature at 8ky b2k
	is $\approx 3$ K higher than the temperature at 225y b2k, in line with \cite{DahlJensen1998, Johnsen1995a,
	Johnsen2001} and \cite{Vinther2009}.
	The first three studies are based on the borehole
	temperature reconstruction while in \cite{Vinther2009} the \delOx signal of the NorthGRIP site
	is corrected for elevation effects assuming a common climate signal between NorthGRIP
	and the Renland icecap - the latter proposed to have experienced minimal changes in elevation
	throughout the Holocene.
	A 3 K Holocene optimum signal is in a broad sense consistent with terrestrial and marine proxy reconstructions
	of temperature for Greenland (Table 3 in \cite{Kaufman2004}).

	In order to obtain a 3 K Holocene optimum signal we assume a linearly varying
	thinning function with respect to depth and set $S'\left( z = 2100 \right) = 0.28$.
	The temperature reconstruction that utilizes the proposed
	modified thinning function is presented in Fig. \ref{Fig16}.
	By adopting this correction we do not impose any change in the higher frequencies of the temperature signal.
	As a result the magnitudes of the BA and the YD transitions remain unaltered.
	The same
	applies for the 8.5k cooling event as well as the rest of the observed Holocene variability.
	Additionally,
	the thinning correction has an impact on the inferred temperature of the Younger -- Dryas stadial, resulting in
	a temperature of $227.5 \pm 2.7 \; \mathrm{K}$.
	With this updated estimate, we achieve an agreement well
	within the $1\sigma$ uncertainty when compared to \cite{Severinghaus1998}.

	Here we assume that the use of the modified thinning function $S'(z)$
	should not imply a change on the layer
	counted GICC05 chronology and the profile of annual layer thickness $\lambda (z)$.
	Hence, the proposed change requires a reduction of the inferred accumulation
	rates.
	In Fig. \ref{Fig18}, we present the old and proposed scenarios for the thinning function and accumulation
	rates. We use the primed notation for the proposed scenarios as $S'(z)$ for the thinning function
	and $A'(z)$ for the accumulation rate.
	The required reduction of the accumulation rate is
	$\approx 10\%$  at 8ky b2k and $\approx 15 \%$ at the time of the Younger--Dryas stadial.

	A similar correction in accumulation rate for the NorthGRIP site has been proposed for
	Marine Isotope Stage 3 (MIS--3) by \cite{Huber2006}.
	A reduction by  $\approx \; 25\%$
	is required in order to eliminate the discrepancy between the output of a firn densification/gas
	fractionation model and \delN measurements over the sequence of Interstadials 9--17.
	Similarly, \cite{Guillevic2013} require a 30\% reduction of past accumulation rates as derived by
	a D--J ice flow model for the NEEM site tuned to the GICC05 chronology on a gas
	fractionation study focusing on Interstadials 8--10.
	In combination with these results,
	our study indicates an inadequacy of the D--J ice flow model in accurately reconstructing
	the strain and accumulation rate histories of the NorthGRIP and NEEM sites. In a synthesis of their
	results with the  study of \cite{Landais2004, Landais2005} , \cite{Guillevic2013} propose that
	this possible inadequacy applies only for glacial conditions, but not for interglacial
	periods as well as glacial inceptions.
	In contrast, our study demonstrates how
	the D--J model fails to accurately estimate the accumulation rate history for the time span
	of the Holocene epoch with an error that is increasing as one approaches glacial conditions.
	Preliminary results, indicate that this error converges to a value of approximately 25\% at
	a depth of 2100 m ($\approx40.2$ky b2k). However, due to the ice diffusion length
	increasing with depth, inference of past temperatures below the depth of $\approx 2000 \mathrm{m}$
	comes with much greater uncertainty.

	One of the possible explanations for the observed error is the use of a constant value used for the
	kink height $\mathcal{H}$ in the D--J ice flow model.
	This is the height that defines the shape
	of the ice horizontal velocity $v_x \left( z \right)$.
	Above $\mathcal{H}$, $v_x$ is assumed to be constant
	and independent of z. Below $\mathcal{H}$, $v_x$ is assumed to linearly decrease until the value of basal
	sliding velocity \citep{Dansgaard1969}.
	The choice of a constant value for $\mathcal{H}$ constitutes
	an oversimplification.
	Measurements of inclination in the bore--holes of the Dye--3 and Camp Century
	sites, reveal an enhanced deformation of the ice below
	$\mathcal{H}$ \citep{Hansen1988, Gundestrup1993}.
	The estimated horizontal velocity profiles from those measurements demonstrate two things.
	First, an apparent kink in the estimated $v_x$ profile
	is observed at the transition between the Holocene and  glacial ice.
	Second, $v_x$ deviates from the assumed constant value already above $\mathcal{H}$ (Fig. 7 and 8 in
	\cite{Hansen1988} and \cite{Gundestrup1993}) respectively.
	These two observations support the theory
	of a moving  kink as also suggested by \cite{Guillevic2013}.
	It is possible that the kink moves along with the interface
	between Holocene and glacial ice owing to the strong gradient in shear strain rates between the two materials.
	The latter is shown to be dependent on the ice fabric, impurity content and grain size of the ice
	\citep{Paterson1991, Thorsteinsson1999, Cuffey2000a}.
	Additionally a concern can be raised regarding the validity
	of the assumption of the constant horizontal velocity from the kink height to the surface.
	We believe that implementations of  ice flow models that take into account the aforementioned
	issues can likely provide better estimates of the ice thinning function and consequently
	past accumulation rates.

	Furthermore, an additional effect that may  have an impact
	on the accuracy of the thinning function estimation by means of a D--J model
	concerns possible migrations of the ice divide in the past.
	Previous studies have
	demonstrated that such effects have occurred at the Siple Dome \citep{Nereson1998}
	and the Greenland summit \citep{Marshall2000}  ice core sites.
	Despite the uncertainty involved in the modeling efforts addressing the problem of
	migrating ice divides, it is safe to assume that the Greenland Ice Sheet was very different
	during glacial times.
	For that reason one can expect that a simple  ice flow model
	is very likely to present inaccuracies   regarding the calculated strain rate history of the
	NorthGRIP site.

\subsection{On the Holocene climate variability}

\begin{figure*}
\vspace*{2mm}
\center
\includegraphics[width= 120mm]{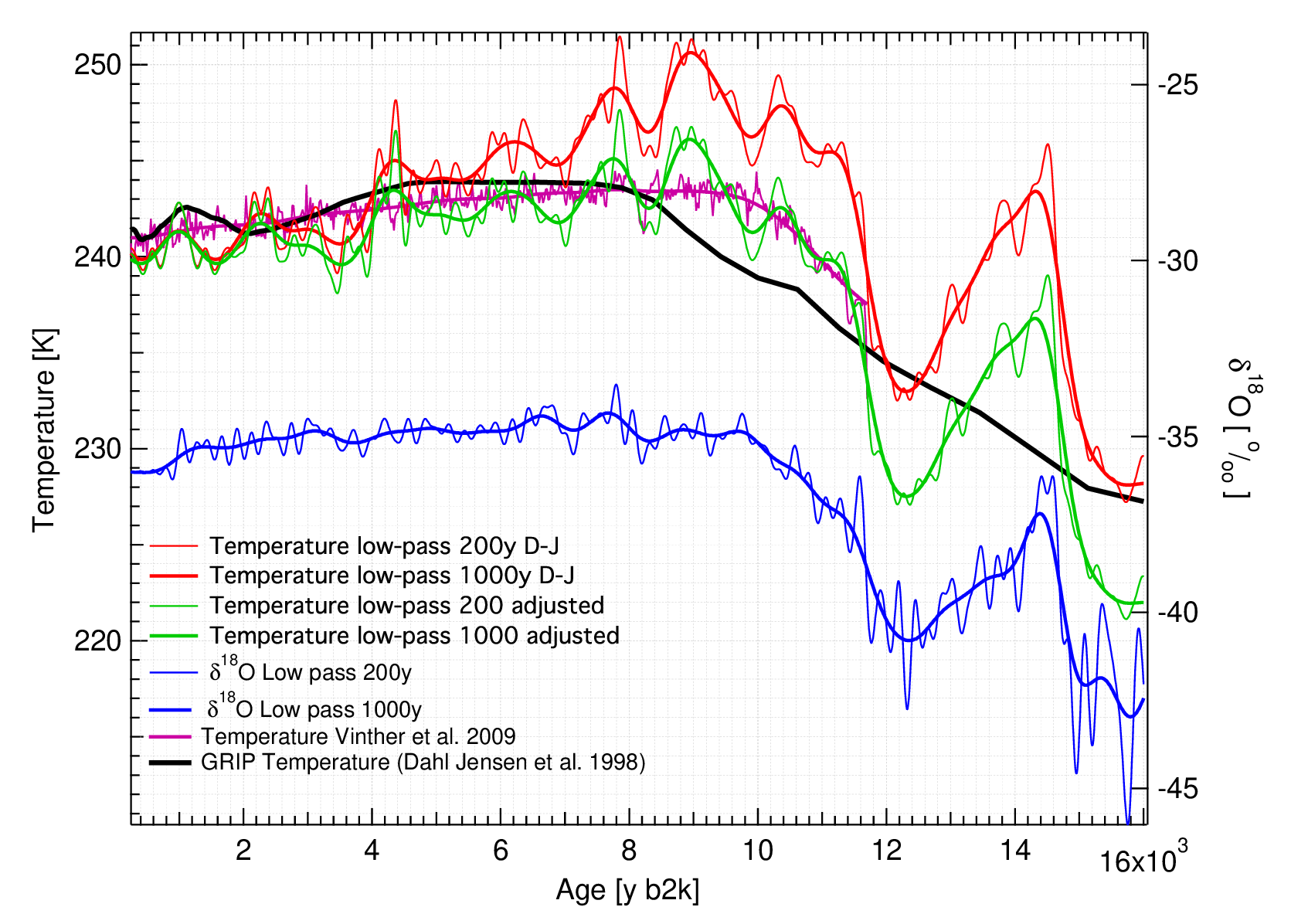}
\caption{The old (red) and updated (green) temperature reconstruction using the proposed thinning
function $S'(z)$. The signal is filtered by applying a low pass filter with a cut--off at 1ky (thick lines) and 200y (thin lines).
In blue we present the low--pass filtered \delOx signal with a cut--off at 1ky (thick line) and 200y (thin line)
In purple we plot the temperature reconstruction from \cite{Vinther2009} where an offset of 241K has been
added and in black the borehole temperature reconstruction by \cite{DahlJensen1998}.}
\label{Fig16}
\end{figure*}

\begin{figure*}
\vspace*{2mm}
\center
\includegraphics[width= 120mm]{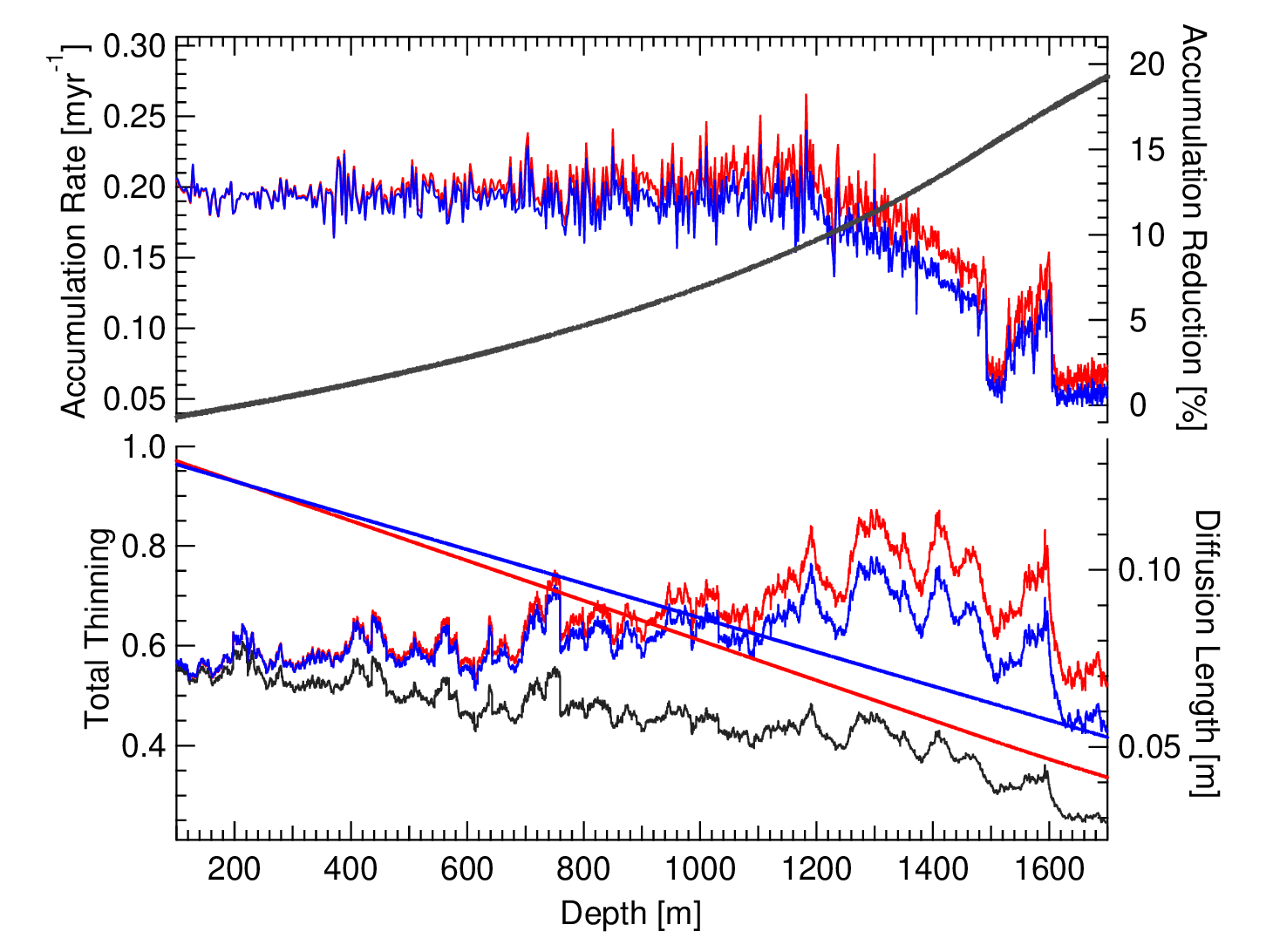}
\caption{The old (red) and new proposed (blue) accumulation rate ($A(z), \;A'(z))$ and thinning history
($S(z), \; S'(z)$) for the North
GRIP site based on the tuning
of the Holocene optimum signal at 8ky b2k to 3 K with respect to the level of 225 y b2k. With the dark gray line we
illustrate the \% proposed reduction in accumulation rates. At the bottom graph the black curve
presents the raw diffusion length signal $\sigma_i\left( z \right)$. Accounting for the two different scenarios of the thinning
function $S\left( z \right)$ and $S'\left( z \right)$ we obtain the red (old) and blue (updated) curve for $\sigma_{\mathrm{firn}} \left( z \right)$ and
$\sigma'_{\mathrm{firn}} \left( z \right)$}
\label{Fig18}
\end{figure*}

	An interesting feature of the diffusion derived temperature history is the observed
	Holocene variability.
	The 200y and 1ky low--pass filtered temperature signals
	in Fig. \ref{Fig16} reveal a climate variability on both millennial and centennial time scales.
	It can be seen in the 1 ky filtered curve (dark blue in Fig. \ref{Fig16}), that the
	cooling trend that started after the Holocene optimum persists through
	the Holocene and it intensifies at around 4 ky b2k, signifying a shift to cooler
	temperatures. In fact, a look into the 200 y low-pass filtered curve
	(light blue in Fig. \ref{Fig16}) reveals a warm event  preceding the shift
	to cooler temperatures and giving rise to a mid-Holocene optimum.
	The climatic shift at around 4ky b2k is of particular interest.
	It possibly indicates that traces of the 4.2ky climatic event,
	observed in a wide range of sites
	\citep{Staubwasser2006, Fisher2008, Berkelhammer2012, Walker2012}
	at low mid and high latitudes, can also be found on the Greenland Ice Sheet.

	An analysis of the derived Holocene climate variability based on the firn
	diffusion reconstruction is not a goal of this study. The several observable
	features of the temperature history possibly resemble climatic events
	documented by other proxies. The reconstruction is broadly similar to
	the borehole temperature reconstruction from GRIP \citep{DahlJensen1998}.
	On the other hand features of the inferred millennial and especially centennial temperature
	variability are inconsistent with the elevation corrected \delOx temperature
	reconstruction from \cite{Vinther2009}
	with parts of the record presenting  high magnitude and changes that are fast.
	This calls for more work on the diffusion reconstruction technique and
	points to the necessity for replication of the results on samples from neighboring ice coring sites.
	The recently drilled and currently analyzed at a high resolution NEEM  ice core \citep{DahlJensen2013}
	can constitute an excellent dataset for such a study.
	Dual \delOx and \delD measurements on this high resolution sample set will also
	allow for a further investigation and application of the differential diffusion method as introduced
	in \cite{Johnsen2000} and applied in the works of  \cite{Simonsen2011} and \cite{Gkinis2011a}.

\section{Conclusions and outlook}
\label{Conclusions}

	Based on a high resolution \delOx record from NorthGRIP, Greenland we estimated the diffusion
	length history for the site. By using a coupled water isotope diffusion and firn densification
	model, in combination with the GICC05 chronology and an estimate of the total thinning function
	obtained with a D--J ice flow model, we reconstructed a temperature history for the period
	400--16,000y b2k. We outlined the technicalities regarding the  implementation of the model
	as well as the spectral based estimation of the diffusion length from the high resolution record.
	Based on a sensitivity study we estimated the $1\sigma$ uncertainty of the reconstructed temperature to be
	approximately 2.7 K.

	The inferred temperature history shows the climatic transition from the Last Glacial to the Holocene
	Optimum and the Holocene cooling trend thereafter. The BA and YD climatic transitions
	are present in the record, with a timing and amplitude that is broadly consistent with previous estimates
	obtained with gas isotopic fractionation methods.
	A cooling event with a duration of approximately 500
	years and an amplitude of 5 K is observed in the early Holocene.
	It marks a possible relationship with signals of similar nature
	from other north high latitude as well as monsoon sites
	at approximately 8.5 ky b2k.

	An overestimation of the temperature of the Holocene Climatic Optimum points to the
	necessity for a correction of the total thinning function. Assuming that the layer
	counted chronology is accurate, this correction would require a reduction of the accumulation rate
	of the order of 10\% at 8ky b2k with respect to the level estimated with the D--J ice flow model.
	We discussed this result with respect to recent \citep{Guillevic2013}  as well as earlier \citep{Huber2006}
	findings based on \delN studies that propose a similar reduction for Interstadials 9--17
	and MIS--3 from the NEEM and NorthGRIP ice core respectively.
	We also  proposed possible
	reasons for this discrepancy related to the constant value of the kink height in the D--J model as
	well as probable past ice divide migrations.

	We believe that the climate variability of our record, spanning millennial to centennial scales,
	is of particular interest and therefore requires further analysis, possibly additional
	evidence from other ice cores from the Greenland Ice Sheet and obviously careful comparisons
	with reconstructions based on other proxies from low, mid and especially high latitudes.
	The NEEM  ice core can potentially provide an excellent record,
	especially when considering the application of recently developed continuous infrared
	techniques \citep{Gkinis2011}  that provide high resolution and precision \delD and \delOx
	datasets. Extending the diffusion reconstructions to the sections of the ice core shallower than the close--off
	depth will also allow for a very useful comparison with the instrumental temperature record.
	Such an attempt will require slight modifications to the diffusion model in order for it to deal
	with the firn layers in which the diffusion process is still under way.

	Our future study objectives will revolve around these topics,
	not only within the scope of studying the past climate during the Holocene epoch, but also  within
	the scope of understanding the mechanisms that affect the responsivity of the \delOx
	signal to a set of input climatic parameters, most notably temperature and furthermore
	result in a low (almost absent) signal to noise ratio during a period that has likely seen
	sizable climatic variations.

\section*{Acknowledgements}
		IRMS  analysis was performed by Anita Boas.
	We would like to thank Eric Steig, Christo Buizert,
	Takuro Kobashi, Bradley Markle and Spruce Schoenemann
	for fruitful discussions.
	We acknowledge funding through NSF
	Award ARC 0806387 and the Lundbek foundation.
	The NorthGRIP project is directed and organized by the
	Department of Geophysics at the Niels Bohr Institute for
	Astronomy, Physics and Geophysics, University of Copenhagen.
	It was being supported by the National Science Foundations of
	Denmark, Belgium, France, Germany, Iceland, Japan, Sweden,
	Switzerland and the United States of America.
	We would like to thank three anonymous reviewers
	for looking into our work in detail and providing thoughtful
	comments and corrections.
	This work would not have been possible without the insight and
	innovative mind of Sigfus J. Johnsen.

\newpage

\bibliographystyle{harvard}


\newpage

\end{document}